\documentstyle[psfig,twocolumn,prl,aps,amssymb]{revtex}
\begin{document}
\wideabs{
\draft

\date{\today} 
\title{Spontaneous stripe order at certain even-denominator fractions in the lowest Landau level}
\author{Seung-Yeop Lee, Vito W. Scarola, and J.K. Jain}
\address{Department of Physics, 104 Davey Laboratory, The Pennsylvania State 
University, University Park, Pennsylvania 16802}
\maketitle

\begin{abstract}

An understanding of the physics of half or quarter filled lowest Landau level has been 
achieved in terms of a Fermi sea of composite fermions,  
but the nature of the state at other even-denominator fractions has remained unclear.
We investigate in this work Landau level fillings  
of the form $\nu=(2n+1)/(4n+4)$, which correspond to
composite fermion fillings $\nu^*=n+1/2$.  By considering
various plausible candidate states, we rule out the composite-fermion Fermi sea and paired 
composite-fermion state at these filling factors, and predict that the system phase separates into  
stripes of $n$ and $n+1$ filled Landau levels of composite fermions.

\end{abstract}

}

The Coulomb interaction between electrons in two dimensions confined to the 
lowest Landau level expresses itself most strongly through the binding of an even number
of quantum mechanical vortices on each electron and thereby creating particles known as 
composite fermions \cite{review,Jain}.
The residual interaction between composite fermions is weak and unimportant  
under many situations, in the sense that it does not alter the nature of the state in a
qualitative manner.  For example, at the odd-denominator fractions $\nu=n/(2pn\pm 1)$, which
correspond to integral fillings $\nu^*=n$ of composite fermions carrying $2p$ vortices, 
the inter-composite fermion (CF) interaction can often be neglected due to the presence 
of the effective cyclotron
gap.  The model of free composite fermions is thus adequate for understanding the 
basic phenomenology of the fractional quantum Hall effect\cite{Tsui} (FQHE). 
In recent years, focus has shifted to the more subtle physics arising from the weak inter-CF
interaction.  It has been shown that it destabilizes delicate FQHE states in higher Landau levels
and at small filling factors in the lowest Landau levels\cite{Kamilla}, causes a pairing instability 
of the CF Fermi sea in the
second Landau level \cite{Scarola} (LL), enters in the spin physics of composite
fermions \cite{Murthy}, and also affects various response functions \cite{Halperin}.

The state at $\nu=1/2p$, obtained in the $n\rightarrow \infty$ limit of the $\nu=n/(2pn\pm 1)$
sequence, is well described
as a Fermi sea of composite fermions carrying $2p$ vortices \cite{half}, called $^{2p}$CFs.
However, our understanding of other even denominator fractions is less satisfactory.
This issue has gained renewed urgency in view
of recent experiments at ultra low temperatures where some new structure has
started to appear {\em between} the $\nu=n/(2pn\pm 1)$ fractions \cite{Pan}.
We will consider in this work even denominator fractions of the form $\nu=(2n+1)/(4n+4)$, which 
correspond to composite fermion filling factors $\nu^*=n+1/2$, according to the relation 
$\nu=\nu^*/(2\nu^*+1)$.
At these filling factors, the topmost CF-Landau level is only partially occupied, and 
it is crucial to incorporate the inter-CF interaction, without which the ground state would have an
enormous degeneracy.  We ask what finer structure the weak interaction produces.
It will be assumed that the spin is frozen, as appropriate for sufficiently high magnetic fields.
Our conclusions will also apply to $\nu=(2n+3)/(4n+4)$ due to particle hole symmetry in the lowest LL.

Several interesting states have been discovered at {\em electron} filling factors 
$\nu=n+1/2$, which will serve as the paradigms for the discussion below. In the lowest Landau level
($\nu=1/2$) electrons transform into composite fermions which condense into  
a $^2$CF Fermi sea \cite{Halperin,half}.  In the second LL ($\nu=2+1/2$) electrons turn into composite 
fermions, which are believed to form Cooper pairs \cite{Scarola,Pan,Willett,Moore}.  
In yet higher Landau levels, electrons do not capture vortices but instead exhibit  
a stripe phase \cite{Koulakov,anisotropy}.

\begin{figure}
\vspace{-0.5cm}
\centerline{\psfig{figure=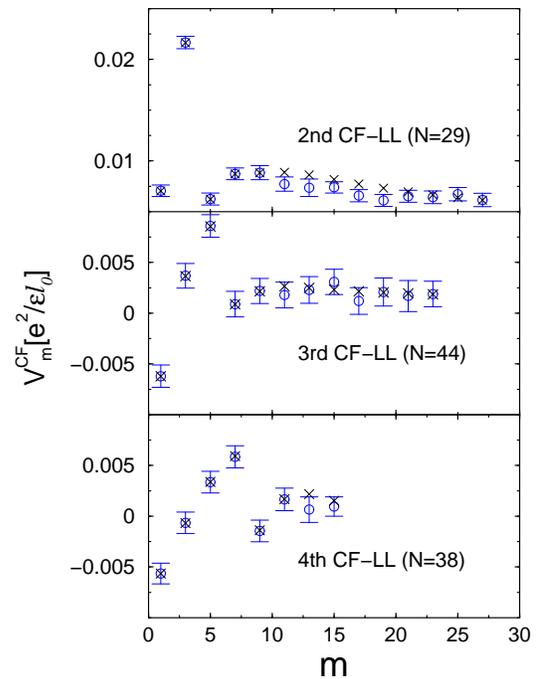,width=3.0in,angle=0}}
\caption{The pseudopotentials 
for the inter-composite fermion interaction in the second, third, and fourth CF Landau levels, 
calculated from the microscopic wave functions (circles).  The error bars indicate the statistical 
error from the Monte Carlo sampling.  The crosses are the pseudopotentials for  
the model interaction $V^{CF}(r)$ explained in the text.} 
\label{fig1}
\end{figure}

Which state actually occurs depends on the interaction matrix elements, and it is therefore 
important to have a good model for the inter-CF interaction, $V^{CF}(r)$.  
We proceed as follows.  In order to
treat the state at $\nu^*=n+1/2$, we 
start with the state with $\nu^*=n$ filled Landau levels of composite fermions and add two additional
composite fermions in the lowest empty CF Landau level.  Following the standard procedure for writing
the wave functions for composite fermions \cite{Jain}, the wave function for this state
is given by ${\cal P}_{LLL}\Phi_1^2 \Phi_n^{++m}$, where 
where $\Phi_1$ is the wave function of one filled LL, $\Phi_n^{++m}$ is the wave function of the {\em
electron} state in which $n$ LLs are fully occupied and the $(n+1)$st LL contains 
two electrons in a relative angular
momentum $m$ state, and ${\cal P}_{LLL}$ is the lowest LL projection operator.  
The explicit form for the general wave functions of this kind is given in the literature \cite{JK}; 
the calculation of energy requires evaluation of  
multi-dimensional integrals which is accomplished by the Monte Carlo method. (The spherical 
geometry \cite{Haldane} is used in our calculations.)  This provides the 
pseudopotentials \cite{Haldane} $V^{CF}_m$, which completely specify the interaction between two
composite fermions in the $(n+1)$st CF-LL.  Similar studies have been done previously \cite{Quinn}, 
except that here $V^{CF}_m$ are evaluated for fairly large systems, 
believed to give a good approximation for the thermodynamic limit.  We then construct a real space
interaction between composite fermions; for convenience, we map the problem of
composite fermions in any arbitrary CF-LL into a
problem of fermions in the lowest LL.  There is no unique prescription 
for this, because many real space interactions produce the same pseudopotentials,
but we find it convenient to use a potential of the form\cite{Park}:
\begin{equation}
V^{CF}(r)=\left(\sum_j c_j r^{2j}e^{-r^2}+
\frac{(2n+1)^{-5/2}}{r}\right)\left[\frac{e^2}{\epsilon l_0}\right]
\end{equation}
The last term gives the Coulomb interaction between two particles of appropriate fractional charge.
The distance $r$ is measured in units of the {\em effective} magnetic length $\ell$, but the 
energies are measured
in units of $e^2/\epsilon l_0$ where $l_0$ is the magnetic length at the actual electron filling
factor $\nu$ and $\epsilon$ is the dielectric constant of the background material.  
We fix the first few parameters $c_j$ by requiring that $V^{CF}(r)$ 
produce the first 5 to 6  odd pseudopotentials exactly.  
A comparison between $V^{CF}_m$ and the pseudopotentials of $V^{CF}(r)$  
(Fig.~\ref{fig1}) shows that $V^{CF}(r)$ is a good approximation for all distances.
We have thus mapped the problem of $N$ composite fermions in the $(n+1)$st LL into that of $N$
fermions at an effective filling in
the lowest LL interacting with an effective potential. 
Only the composite fermions in the topmost half filled CF-LLs will be considered explicitly; the 
completely occupied CF-LLs appear only through their role 
in determining the inter-CF interaction.
We note that the interaction between composite fermions is remarkedly different from 
that between electrons in the corresponding higher LLs.
In the second CF-LL, it is most strongly repulsive in the $m=3$ channel, and in higher
CF-LLs, the interaction is actually attractive, with the lowest energy obtained in the $m=1$ channel.

Our conclusions below will be subject to two assumptions.  (i) 
We assume that mixing with higher CF-LLs can be neglected, i.e., the inter-CF
interaction is weak compared to the effective CF-cyclotron energy.  There
is evidence that this
is an excellent approximation: the states containing several composite fermions are accurately
described without considering mixing between CF-LLs \cite{Dev}.
(ii) We further assume that the interaction energy of many 
composite fermions in the $(n+1)$st CF-LL is well approximated by
a sum of two-body terms.  An examination of configurations containing three composite
fermions in the second CF-LL indicates that this is a good approximation \cite{Quinn,Sing}.

\begin{figure}
\vspace{-.5cm}
\centerline{\psfig{figure=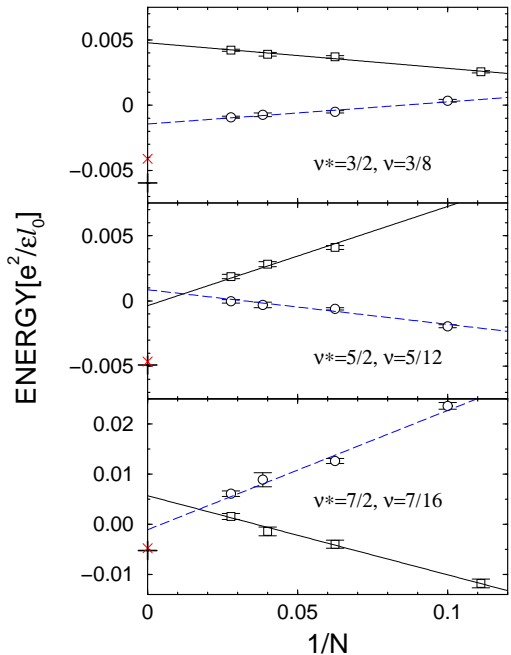,width=3.0in,angle=0}}
\caption{The energy per particle for the CF Fermi sea (squares)
and the CF paired state (circles) as a function of $N$, the number of composite fermions in the 
$(n+1)$st CF-Landau level.  The thermodynamic energy is also shown for the CF 
stripe and bubble phases (dash and cross, respectively). All energies are 
measured relative to the uncorrelated uniform density state, explained in the text.}
\label{fig2}
\end{figure}

The first state that we consider is the Fermi sea, in which the $^2$CFs capture two 
additional vortices to become $^4$CFs, which then form a Fermi sea. (The composite fermions 
in the lower, fully occupied CF-LLs remain
$^2$CFs; this state thus contains an admixture of two different 
flavors of composite fermions.)  The wave function of the Fermi sea is 
\begin{equation}
\Psi_{FS}={\cal
P}_{LLL}\Phi_1^2\Phi_{\infty}
\end{equation} 
The thermodynamic limit for the energy of the $^4$CF sea is
obtained by an extrapolation of finite system results, as shown in Fig.~(\ref{fig2}).  
Since we are interested in comparing energies obtained by 
different methods, it is important to carefully define the total energy; we will consistently  
take the same form for the electron-electron, electron-background, and background-background 
interactions in all our calculations.  
All energies are quoted relative to the energy of an uncorrelated uniform 
state, explained below.

We do not expect the ground state to be the $^4$CF sea because 
the stability of the Fermi sea requires a strong short range
repulsion, which is not the case with composite fermions in higher CF-LLs \cite{footnote}.
In fact, in the third and higher CF-LLs, the interaction between the composite 
fermions is attractive, which might suggest pairing of composite fermions.  
The paired state of composite fermions is represented by the Pfaffian wave 
function\cite{Moore}
\begin{eqnarray}
\Psi_{Pf} = \Phi_1^2\; {\rm Pf} [M]
\end{eqnarray}
where ${\rm Pf}[M]$ is the Pfaffian of the $N \times N$ antisymmetric
matrix $M$ with components $M_{ij} = (u_i v_j - v_i u_j)^{-1}$, where $u_j\equiv
\cos(\theta_j/2)\exp(-i\phi_j/2)$
and $v_j\equiv \sin(\theta_j/2)\exp(i\phi_j/2)$. 
${\rm Pf} [M]$ is a real space BCS wave function, so $\Psi_{Pf}$ describes 
a paired state of composite fermions.
Again, since our base particle is a $^2$CF, $\Psi_{Pf}$ contains pairing of $^4$CFs.
The energy of this state, given in Fig.~(\ref{fig2}), beats the Fermi 
sea at $\nu=3/8$ and $7/16$, raising the intriguing possibility of a FQHE, induced 
by pairing, at certain even denominator 
fractions in the lowest Landau level.

\begin{figure}
\centerline{\psfig{figure=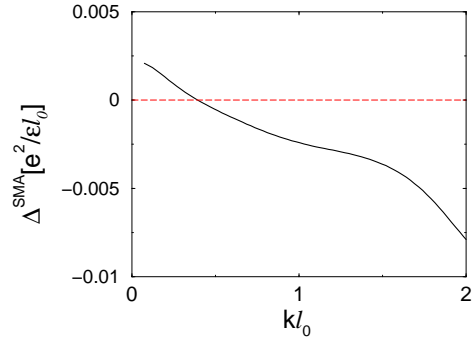,height=2.0in,angle=-90}}
\caption{The excitation energy of the single mode approximation (SMA) excitation  
for the Pfaffian wave function.}
\label{fig3}
\end{figure}

However, it is important to study the stability of any candidate 
FQHE state to quantum fluctuations \cite{Kamilla}. We consider the density-wave 
excitation of the Pfaffian wave function in the single mode approximation \cite{GMP}, 
described by the wave function 
${\cal P}_{LLL}\rho_k\Psi_{Pf}$,  where $\rho_k$ is the density operator at wave vector $k$.  We  
calculate the energy of this excitation following Ref.~\onlinecite{GMP}, with the help of the
pair correlation function of $\Psi_{Pf}$ \cite{Park2}.  The excitation energy is  
shown in Fig.~(\ref{fig3}) as a function of the wave vector, and indicates that the paired 
state is unstable.

These results rule out the $^4$CF
Fermi sea as well as $^4$CF pairing at the filling factors considered here.  
We have also carried out \cite{Sing} exact
diagonalization at the flux values corresponding to the Fermi sea and the 
Pfaffian wave function in the spherical geometry, and found that the ground
state does not have orbital angular momentum $L=0$, i.e., is not a uniform density state.  
The fact that the instability occurs at 
non-zero wave vectors in Fig.~(\ref{fig3}) also hints that the true 
ground state may not be a translationally invariant liquid.

Besides pairing, another possible consequence of an attractive interaction is phase separation.  
Due to the long range Coulomb interaction, the phase separation is likely to 
manifest itself here through the formation of stripes.
We calculate the Hartree-Fock energy of the stripe state
of $^2$CFs following the method of Koulakov {\em et
al.}\cite{Koulakov}.  The interaction Hamiltonian is given by  
\begin{equation}
{\hat V}=\frac{(2\pi)^3}{2L_xL_y}\sum_{\bf q} \tilde V^{CF}({\bf q})\rho(-{\bf q})\rho({\bf q})
\end{equation}
where $\rho({\bf q})$ is the density operator and $\tilde V^{CF}({\bf q})$ is the Fourier transform of
$V^{CF}(r)$.  
Subsequent to 
a Hartree-Fock decomposition, the expectation value of the interaction energy can be written as
\begin{equation}
<{\hat V}>=\frac{(2\pi)^3}{2L_xL_y}\sum_{\bf q}\tilde U_{HF}(q)\Delta(-{\bf
q})\Delta({\bf q})
\end{equation}
where $\Delta({\bf q})\equiv\frac{1}{2\pi}\sum\nolimits_ke^{-ikq_x\ell^2}<a^{\dagger}_{k_+}a_{k_-}>$, 
$\tilde U_{HF}(q)=\tilde U(q)-\ell^2 U(q\ell^2)$, 
$\tilde U(q)\equiv V^{CF}(q)e^{-\frac{1}{2}q^2\ell^2}$, and $k_{\pm} =k\mp q_y/2$.
The ${\bf q}=0$ term, which corresponds to the uniform (uncorrelated) state,
is treated separately:  the direct part is canceled by the background, and the
exchange part is taken as the reference energy, given by 
\begin{equation}
E_{0}=-N{\overline \nu}\frac{U(0)}{2}
\end{equation}
where ${\overline \nu}=1/2$.
The contribution coming from nonzero values of ${\bf q}$ is called the coherence energy, $E_{coh}$.

The stripe phase with period $\Lambda$ corresponds to the choice:
\begin{equation}
\Delta(x,y)=\frac{1}{2\pi\ell^2}\sum_q \frac{2\sin(\frac{q\Lambda{\overline\nu}}{2})}{\Lambda q}e^{iqx}
\end{equation}
where $q=\frac{2j \pi}{\Lambda}$, with $j=\pm 1, \pm 2, \cdots$.
This gives 
\begin{equation}
E_{coh}=\frac{1}{2{\overline \nu}\ell^2}\sum_q\tilde U_{HF}(q)\left(\frac{2\sin(\frac{q\Lambda{\overline \nu}}
{2})}{\Lambda q}\right)^2
\end{equation}
We compute it as a function of $\Lambda$.
The lowest energy, shown in Fig.~(\ref{fig2}),
is obtained at $\Lambda/l_0=$ 10, 28, and 34 for $\nu=3/8$, 5/12, and 7/16.
The period is rather large compared to that for the electron stripes in higher LLs 
(for which $\Lambda/l_0$ is of order unity), which is not surprising because
(i) the period of the stripes is controlled by the effective magnetic length, and (ii) 
the difference between the densities of the FQHE states on either side is quite small.
The stripe phase has lower energy than both the Fermi-sea and paired states.

For completeness, we have also considered the Wigner crystal of ``bubbles" \cite{Koulakov}, with 
each bubble containing in general several electrons.  
For a honeycomb lattice with lattice constant $\Lambda_b$, the radius 
of a bubble is $R=\sqrt{\frac{\sqrt{3}{\overline \nu}}{2\pi}}\Lambda_b$
and 
\begin{equation}
\Delta({\bf r})=
\frac{2}{\sqrt{3}\Lambda_b^2}\sum_{\bf q}\frac{R}{\ell^2 q}J_1(qR)e^{i{\bf q}\cdot{\bf r}}
\end{equation}
where ${\bf q}=m_1{\bf e_1}+m_2{\bf e_2}$ with ${\bf e_1}=\frac{4\pi}{\sqrt{3}\Lambda_b}\hat y$, 
${\bf e_2}=\frac{2\pi}{\Lambda_b}\hat x-\frac{2\pi}{\sqrt{3}\Lambda_b}\hat y$, and $m_1$ and $m_2$
being integers.  The coherence energy for the bubble phase is 
\begin{equation}
E_{coh}=\frac{4\pi}{\sqrt{3}\ell^2\Lambda^2_b}\sum_{\bf q}\tilde U_{HF}(q)\left(\frac{R}{A\ell^2
q}\,J_1(qR)\right)^2
\end{equation}
The lowest energy, shown in Fig.~(\ref{fig2}), is determined by considering 
bubbles with various occupancies, and has higher energy than the stripe phase 
at all of the filling factors considered here.

In the above, the filling factor $\nu^*=n+1/2$ is viewed as half filling of CF particles on top of
$n$ filled CF-LLs.  We have also considered\cite{Sing} the complementary approach in which it 
is modeled as half filling of CF holes on the background of 
$n+1$ filled CF-LLs.   The stripe phase again has the lowest
energy, which gives us further confidence in the robustness of our result.

The electron stripes in higher electronic Landau levels have revealed themselves through an
anisotropic transport at temperatures below $\sim $ 50 mK \cite{anisotropy}.  The conditions for the
observation of CF stripes are more stringent.  Estimates of the
critical temperature from the Hartree Fock theory are not quantitatively reliable, but noting that
the effective interaction between composite fermions at $\nu^*=n+1/2$ is 
roughly an order of magnitude smaller than for electrons at $\nu=n+1/2$, 
as measured by the pseudopotentials, 
we expect the critical temperature to also be similarly reduced.
Also, the much larger period suggests the need for a high degree of density homogeneity.

This work was supported in part by the National Science
Foundation under grants DMR-9986806 and DGE 9987589.

\end{document}